# Key Considerations for the Responsible Development and Fielding of Artificial Intelligence

Eric Horvitz     Jessica Young     Rama G. Elluru     Chuck Howell

**Abstract**

We review key considerations, practices, and areas for future work aimed at the responsible development and fielding of AI technologies. We describe critical challenges and make recommendations on topics that should be given priority consideration, practices that should be implemented, and policies that should be defined or updated to reflect developments with capabilities and uses of AI technologies. The Key Considerations were developed with a lens for adoption by U.S. government departments and agencies critical to national security. However, they are relevant more generally for the design, construction, and use of AI systems.

April 2021

**Forward**

The paradigm and recommended practices described here stem from the National Security Commission on Artificial Intelligence's (NSCAI) "line of effort" on Ethics and Responsible Artificial Intelligence (AI). The development of the content was supported by input and feedback drawn from consultations with representatives from academia, civil society, industry, and federal agencies. These recommended considerations and practices provide the foundation for concepts on the responsible development and fielding of AI technologies appearing throughout the NSCAI report, including the Commission's recommendations. The NSCAI report was delivered to Congress and the Executive Branch on March 1, 2021. The concepts in this document were instrumental in developing the content appearing in Chapters 7 on "Establishing Justified Confidence in AI Systems," and Chapter 8 on "Upholding Democratic Values: Privacy, Civil Liberties, and Civil Rights in Uses of AI for National Security."

An earlier version of this document was published by the NSCAI as an appendix in the "Second Quarter Recommendations" of the Commission in July 2020. An updated version is included as an appendix in the final NSCAI report, entitled, "Key Considerations for the Responsible Development and Fielding of Artificial Intelligence." The NSCAI final report was approved by all 15 NSCAI Commissioners (Safra Catz, Dr. Steve Chien, Hon. Mignon Clyburn, Chris Darby, Dr. Ken Ford, Dr. José-Marie Griffiths, Andy Jassy, Gilman Louie, Dr. William Mark, Dr. Jason Matheny, Hon. Katharina McFarland, and Dr. Andrew Moore, Dr. Eric Schmidt, Hon. Robert O. Work, in addition to Dr. Eric Horvitz, who chaired the line of effort on ethics and responsible AI). Commissioners Horvitz, Matheny, Clyburn, and Griffiths served on the NSCAI line of effort on Ethics and Responsible AI.

## Introduction

Concerns about the responsible development and fielding of AI technologies span a range of issues. Discussions and debates are ongoing as the technology and its applications rapidly evolve, and the need for norms and best practices becomes more apparent.

Efforts have been undertaken to date to establish ethics guidelines for AI by entities in government, in the private sector, and around the world.[1] In 2019, forty-two countries including the U.S. adopted the OECD Principles on Artificial Intelligence.[2] Within the U.S. Government, the Department of Defense took the critical step of adopting a set of high-level principles to guide its development and use of AI,[3] followed by the Office of the Director of National Intelligence's (ODNI) adoption of AI principles for the Intelligence Community (IC).[4] In 2020, Executive Order 13960 further established Principles for Use of AI in Government.[5] However, even in cases where principles are offered, it can be difficult to translate the high-level concepts into concrete actions. There is often a gap between articulating high-level goals around responsible AI and operationalizing them.

The ideas in this manuscript are based on an assessment of current challenges for responsibly developing and fielding AI systems, and the practices and future directions needed to overcome these challenges. We focus our discussion of these challenges and recommendations within the context of five categories: Aligning AI Systems with Democratic Values and the Rule of Law; Engineering Practices; System Performance; Human-AI Interaction and Teaming; and Accountability and Governance. By assessing key considerations for responsible AI adoption within the context of uses by national security departments and agencies, we explore considerations that are especially imperative in high-stakes deployments, including those of safety-critical systems and those with acute implications for the preservation of life and liberties.

---

[1] Examples of efforts to establish ethics guidelines are found within the U.S. government, industry, and internationally. See e.g., Draft Memorandum for the Heads of Executive Departments and Agencies: Guidance for Regulation of Artificial Intelligence Applications, Office of Management and Budget (Jan. 1, 2019), https://www.whitehouse.gov/wp-content/uploads/2020/01/Draft-OMB-Memo-on-Regulation-of-AI-1-7-19.pdf; Jessica Fjeld & Adam Nagy, Principled Artificial Intelligence: Mapping Consensus in Ethical and Rights-Based Approaches to Principles for AI, Berkman Klein Center (Jan. 15, 2020), https://cyber.harvard.edu/publication/2020/principled-ai; OECD Principles on AI, OECD (last accessed June 17, 2020), https://www.oecd.org/going-digital/ai/principles/; Ethics Guidelines for Trustworthy AI, High-Level Expert Group on Artificial Intelligence, European Union at 26-31 (Apr. 8, 2019), https://ec.europa.eu/futurium/en/ai-alliance-consultation/guidelines.

[2] See https://www.oecd.org/science/forty-two-countries-adopt-new-oecd-principles-on-artificial-intelligence.htm.

[3] C. Todd Lopez, DOD Adopts 5 Principles of Artificial Intelligence Ethics, U.S. Department of Defense (Feb. 5, 2020), https://www.defense.gov/Explore/News/Article/Article/2094085/dod-adopts-5-principles-of-artificialintelligence-ethics/ [hereinafter, Lopez, DOD Adopts 5 Principles].

[4] See Principles of Artificial Intelligence Ethics for the Intelligence Community, Office of the Director of National Intelligence (last accessed Jan. 11, 2021), https://www.dni.gov/index.php/features/2763-principles-of-artificialintelligence-ethics-for-the-intelligence-community.

[5] See Donald J. Trump, Executive Order on Promoting the Use of Trustworthy Artificial Intelligence in the Federal Government, The White House (Dec. 3, 2020), https://trumpwhitehouse.archives.gov/articles/promoting-use-trustworthy-artificial-intelligence-government/.

Section I. of this document provides guidance specific to implementing systems that abide by democratic values and the rule of law. The section covers aligning the run-time behavior of systems to the related, more technical encodings of objectives, utilities, and tradeoffs. The four following sections (on II. Engineering Practices, III. System Performance, IV. Human-AI Interaction, and V. Accountability & Governance) serve in support of core democratic values and outline practices needed to develop and field systems that are trustworthy, understandable, reliable, and robust. Recommended practices span multiple phases of the AI lifecycle, from conception and early design, through development and testing, and maintenance and technical refresh. "Development" refers to 'designing, building, and testing during development and prior to deployment' and "fielding" to refer to 'deployment, monitoring, and sustainment.'

Though best practices will evolve, these recommended practices establish a baseline for the responsible development and fielding of AI technologies. They provide a floor, rather than a ceiling, for the responsible development and fielding of AI technologies.

Within each of the five sections of this manuscript, we first provide a conceptual overview of the scope and importance of the topic. We then illustrate an example of a current challenge relevant to national security departments that underscores the need to adopt recommended practices in this area. Then, we provide a list of recommended practices that should be adopted, acknowledging research, industry tools, and exemplary models within government that could support adoption of recommended practices. Finally, in areas where recommended practices do not exist or they are especially challenging to implement, we note the need for future work as a priority; this includes, for example, R&D and standards development. We also identify potential areas in which collaboration with allies and partners would be beneficial for interoperability and trust, and we note that the Key Considerations can inform potential future efforts to discuss military uses of AI with strategic competitors.

## I. Aligning Systems and Uses with American Values and the Rule of Law

(1) Overview

Our values guide our decisions and our assessment of their outcomes. Our values shape our policies, our sensitivities, and how we balance tradeoffs among competing interests. Our values, and our commitment to upholding them, are reflected in the U.S. Constitution, and our laws, regulations, programs, and processes.

One of the seven principles we set forth in our 2019 Interim Report is the following:

> The American way of AI must reflect American values—including having the rule of law at its core. For federal law enforcement agencies conducting national security investigations in the United States, that means using AI in ways that are consistent with constitutional principles of due process, individual privacy, equal protection, and nondiscrimination. For American diplomacy, that means standing firm against uses of AI by authoritarian governments to repress individual freedom or violate the human rights of their citizens. And for the U.S. military, that means finding ways for AI to enhance its ability to uphold the laws of war and ensuring that current frameworks adequately cover AI.[6]

Values established in the U.S. Constitution, and further operationalized in legislation, include freedoms of speech and assembly, the rights to due process, inclusion, fairness, non-discrimination (including equal protection), and privacy (including protection from unwarranted government interference in one's private affairs).[7] Beyond the values codified in the U.S. Constitution and the U.S. Code, our values also are expressed via international treaties that the United States has ratified that affirm our commitments to human rights and human dignity, including the International Convention of Civil and Political Rights.[8] Within America's national security departments, our commitment to protecting and upholding privacy and civil liberties is further embedded in the policies and programs of the IC,[9] the Department of Homeland Security (DHS),[10] the Department of Defense (DoD),[11] and oversight entities.[12] This is not an exhaustive set of values that U.S. citizens would identify as core principles of the United States. However, the paradigm of considerations and recommended practices for AI that we introduce resonate with these highlighted values as they have been acknowledged and elevated as critical by the U.S. government and national security departments and agencies. Further, many of these values are common to America's like-minded partners who share a commitment to democracy, human dignity, and human rights.

---

[6] 2019 Interim Report at 17, https://www.nscai.gov/wp-content/uploads/2021/01/NSCAI-Interim-Report-for-Congress_201911.pdf [hereinafter Interim Report].

[7] See e.g., U.S. Const. amendments I, IV, V, and XIV; Americans with Disability Act of 1990, 42 U.S.C. § 12101 et seq.; Title VII of the Consumer Credit Protection Act, 15 U.S.C. §§ 1691-1691f; Title VII of the Civil Rights Act of 1964, 42 U.S.C. § 2000e seq.

[8] International Covenant on Civil and Political Rights, UN General Assembly, United Nations, Treaty Series, vol. 999 at 171 (Dec. 16, 1966), https://www.refworld.org/docid/3ae6b3aa0.html. As noted in the Commission's 2019 Interim Report, America and its like-minded partners share a commitment to democracy, human dignity and human rights. See 2019 Interim Report at 48. Many, but not all nations, share commitments to these values. Even when values are shared, however, they can be culturally relative, for instance, across nations, owing to interpretative nuances.

In the military context, core values such as distinction and proportionality are embodied in the nation's commitment to, and the DoD's policies to uphold, the Uniform Code of Military Justice and the Law of Armed Conflict (LOAC).[13] Other values are reflected in treaties, rules, and policies such as the Convention

---

[9] See e.g., Daniel Coats, Intelligence Community Directive 107, ODNI (Feb. 28, 2018), https://fas.org/irp/dni/icd/icd-107.pdf (on protecting civil liberties and privacy); IC Framework for Protecting Civil Liberties and Privacy and Enhancing Transparency Section 702, Intel.gov (Jan. 2020), https://www.intelligence.gov/index.php/ic-on-the-record/guide-to-posted-documents#SECTION_702OVERVIEW (on privacy and civil liberties implication assessments and oversight); Principles of Professional Ethics for the Intelligence Community, ODNI (last visited June 17, 2020), https://www.dni.gov/index.php/whowe-are/organizations/clpt/clpt-related-menus/clpt-related-links/ic-principles-of-professional-ethics (on diversity and inclusion).

[10] See e.g., Privacy Office, U.S. Department of Homeland Security (June 3, 2020), https://www.dhs.gov/privacy-office#; CRCL Compliance Branch, U.S. Department of Homeland Security (May 15, 2020), https://www.dhs.gov/compliance-branch.

[11] See Samuel Jenkins & Alexander Joel, Balancing Privacy and Security: The Role of Privacy and Civil Liberties in the Information Sharing Environment, IAPP Conference 2010 (2010), https://dpcld.defense.gov/Portals/49/Documents/Civil/IAPP.pdf.

[12] See Projects, U.S. Privacy and Civil Liberties Oversight Board, (last visited June 17, 2020), https://www.pclob.gov/Projects.

[13] See Department of Defense Law of War Manual, U.S. Department of Defense (Dec. 2016), https://dod.defense.gov/Portals/1/Documents/pubs/DoD%20Law%20of%20War%20Manual%20-%20June%202015%20Updated%20Dec%202016.pdf?ver=2016-12-13-172036-190 [hereinafter DoD Law of War Manual]. See also AI Principles: Recommendations on the Ethical Use of Artificial Intelligence by the Department of Defense: Supporting Document, DoD Defense Innovation Board (Oct. 31, 2019), https://media.defense.gov/2019/Oct/31/2002204459/-1/-1/0/DIB_AI_PRINCIPLES_SUPPORTING_DOCUMENT.PDF ("More than 10,000 military and civilian lawyers within DoD advise on legal compliance with regard to the entire range of DoD activities, including the Law of War. Military lawyers train DoD personnel on Law of War requirements, for example, by providing additional Law of War instruction prior to a deployment of forces abroad. Lawyers for a Component DoD organization advise on the issuance of plans, policies, regulations, and procedures to ensure consistency with Law of War requirements. Lawyers review the acquisition or procurement of weapons. Lawyers help administer programs to report alleged violations of the Law of War through the chain of command and also advise on investigations into alleged incidents and on accountability actions, such as commanders' decisions to take action under the Uniform Code of Military Justice. Lawyers also advise commanders on Law of War issues during military operations.").

Against Torture and Other Cruel, Inhuman or Degrading Treatment or Punishment;[14] the DoD's Rules of Engagement;[15] and DoD Directive 3000.09.[16]

U.S. values demand that the development and use of AI respect these foundational values, and that they enable human empowerment as well as accountability. They require that the operation of AI systems and components be compliant with our laws and international legal commitments, and with departmental policies. In short, core democratic values must inform the way we develop and field AI systems, and the way our AI systems behave in the world.

To date, AI Principles adopted and endorsed by the U.S. Executive Branch, including by national security department and agencies, have focused on aligning AI with many of the values discussed in this section, including fairness and non-discrimination,[17] privacy and civil liberties,[18] and accountability.[19] Taking the DoD Principles as one example, fairness is evoked by the "Equitable" principle that the Department will "take deliberate steps to minimize unintended bias in AI capabilities."[20] Accountability is evoked by the "Responsible" principle that "DoD personnel will exercise appropriate levels of judgment and care while remaining responsible for the development, deployment and use of AI capabilities."[21] The work on establishing principles reiterates the importance of developing and deploying AI systems in accordance with these values.

---

[14] Convention against Torture and Other Cruel, Inhuman or Degrading Treatment or Punishment, United Nations General Assembly (Dec. 10, 1984), https://www.ohchr.org/en/professionalinterest/pages/cat.aspx.

[15] See DoD Law of War Manual at 26 ("Rules of Engagement reflect legal, policy, and operational considerations, and are consistent with the international law obligations of the United States, including the law of war.").

[16] See Department of Defense Directive 3000.09 on Autonomy in Weapons Systems, U.S. Department of Defense (Nov. 21 2012), https://www.esd.whs.mil/Portals/54/Documents/DD/issuances/dodd/300009p.pdf ("Autonomous and semi-autonomous weapon systems shall be designed to allow commanders and operators to exercise appropriate levels of human judgment over the use of force.").

[17] See e.g., Lopez, DOD Adopts 5 Principles; Draft Memorandum for the Heads of Executive Departments and Agencies: Guidance for Regulation of Artificial Intelligence Applications, Office of Management and Budget (Jan. 1, 2019), https://www.whitehouse.gov/wp-content/uploads/2020/01/Draft-OMB-Memo-on-Regulation-ofAI-1-7-19.pdf.

[18] See Principles of Artificial Intelligence Ethics for the Intelligence Community, ODNI (last accessed Jan. 11, 2021), https://www.dni.gov/index.php/features/2763-principles-of-artificial-intelligence-ethics-for-theintelligence-community.

[19] Id.

[20] See Lopez, DOD Adopts 5 Principles.

[21] Id.

## (2) Examples of Current Challenges

With respect to the U.S. Government, machine learning techniques can assist DoD agencies with conducting large scale data analyses to support and enhance decision-making about personnel. As an example, the Joint Artificial Intelligence Center (JAIC) Warfighter Health Mission Initiative Integrated Disability Evaluation System model seeks to leverage data analyses to identify service members on the verge of ineligibility due to concerns with their readiness.[22] Other potential analyses can support personnel evaluations, including analyzing various factors that lead to success or failure in promotion. Caution and proven practices are needed, however, to avoid pitfalls in fairness and inclusiveness, several of which have been highlighted in high-profile challenges in such areas as criminal justice,[23] recruiting and hiring,[24] and face recognition.[25] Attention should be paid to challenges with decision support systems to avoid harmful disparate impact.[26] Likewise, factors chosen to weigh in performance evaluations and promotions must be carefully considered to avoid inadvertently reinforcing existing biases through ML-assisted decisions.

---

[22] See JAIC Mission Initiative in the Spotlight: Warfighter Health, JAIC (Apr. 15, 2020), https://www.ai.mil/blog_04_15_20-jaic_mi_warfighter_health.html.

[23] Report on Algorithmic Risk Assessment Tools in the U.S. Criminal Justice System, Partnership on AI, (last accessed July 14, 2020), https://www.partnershiponai.org/report-on-machine-learning-in-risk-assessmenttools-in-the-u-s-criminal-justice-system/.

[24] Andi Peng et al., What You See Is What You Get? The Impact of Representation Criteria on Human Bias in Hiring, Proceedings of the 7th AAAI Conference on Human Computation and Crowdsourcing (Oct. 2019), https://arxiv.org/pdf/1909.03567.pdf; Jeffrey Dastin, Amazon Scraps Secret AI Recruiting Tool that Showed Bias Against Women, Reuters (Oct. 9, 2018), https://www.reuters.com/article/us-amazon-com-jobs-automationinsight/amazon-scraps-secret-ai-recruiting-tool-that-showed-bias-against-women-idUSKCN1MK08G [hereinafter Dastin, Amazon Scraps Secret AI Recruiting Tool].

[25] Patrick Grother, et. al., Face Recognition Vendor Test (FRVT) Part Three: Demographic Effects, National Institute of Standards and Technology (Dec. 2019), https://doi.org/10.6028/NIST.IR.8280 [hereinafter Grother, Face Recognition Vendor Test (FRVT) Part Three: Demographic Effects].

[26] PNDC provides predictive analytics to improve military readiness; enable earlier identification of service members with potential unfitting, disabling, or career-ending conditions; and offer opportunities for early medical intervention or referral into disability processing. To do so, PNDC provides recommendations at multiple points in the journey of the non-deployable service member through the Military Health System to make "better decisions" that improve medical outcomes and delivery of health services. This is very similar to the OPTUM decision support system that recommended which patients should get additional intervention to reduce costs. Analysis showed millions of US patients were processed by the system, with substantial disparate impact on black patients compared to white patients. Shaping development from the start to reflect bias issues (which can be subtle) would have produced a more equitable system and avoided scrutiny and suspension of system use when findings were disclosed. See Heidi Ledford, Millions of Black People Affected by Racial Bias in Health Care Algorithms, Nature (Oct. 26, 2019), https://www.nature.com/articles/d41586-01903228-6.

(3) Recommended Practices

A. Developing uses and building systems that behave in accordance with democratic values and the rule of law.

1. **Employ technologies and operational policies that align with privacy preservation, fairness, inclusion, human rights, and law of armed conflict.** Technologies and policies throughout the AI lifecycle should support achieving the goals that AI systems and uses are consistent with these values—and should mitigate the risk that AI system uses/outcomes will violate these values.

   ● An explicit analysis of outcomes that would violate these values should be performed. Policy should prohibit disallowed outcomes that would violate the values above. During system development, analysis of system-specific disallowed outcomes should be performed.[27] As the technology advances, applications evolve, and our understanding of the implications of use grows, these policies should periodically be refreshed.

   ● While not an exhaustive list, we offer the following examples based upon core values discussed above:

     o For ensuring privacy, employ privacy protections, privacy-sensitive analyses, ML with encrypted data and models, and multiparty computation methods. Use

---

[27] This combined approach of stable policy-level disallowed outcomes and system-specific disallowed outcomes is consistent with DoD practices for system safety, for example. See Department of Defense Standard Practice: System Safety, U.S. Department of Defense (May 11, 2012), https://www.dau.edu/cop/armyesoh/DAU%20Sponsored%20Documents/MIL-STD-882E.pdf. Depending on the context, mitigating harm per values and disallowed outcomes might entail the use of fail-safe technologies. See Eric Horvitz, Reflections on Safety and Artificial Intelligence, Exploratory Technical Workshop on Safety and Control for AI (June 27, 2016), http://erichorvitz.com/OSTP-CMU_AI_Safety_framing_talk.pdf. See also Dorsa Sadigh & Ashish Kapoor, Safe Control Under Uncertainty with Probabilistic Signal Temporal Logic, Proceedings of Robotics: Science and Systems XII (2016), https://www.microsoft.com/en-us/research/wpcontent/uploads/2016/11/RSS2016.pdf.

evolving metrics to calibrate risk exposure to privacy attacks[28] and take steps to mitigate such attacks.[29]

- o For fairness and to mitigate unwanted bias, work with stakeholders to define goals of fairness for a system, make the definitions accessible for inspection, and use tools to probe for unwanted bias in data, inferences, and recommendations.[30]

- o For inclusion, ensure usability of systems, accessible design, appropriate ease of use, learnability, and training availability.

- o For commitment to human rights, place limitations and constraints on applications that would put commitment to human rights at risk, for example, limits on storing observational data beyond its specific use or using data for purposes other than its primary, intended focus.

- o For compliance with the Law of Armed Conflict, tools for interpretability and to provide cues to the human operator should enable context-specific judgments to ensure, for instance, distinction between active combatants, those who have surrendered, and civilians.[31]

---

[28] See Nicholas Carlini, et al., The Secret Sharer: Evaluating and Testing Unintended Memorization in Neural Networks, Usenix Security Symposium 2019 (Aug. 14-16, 2019), https://www.semanticscholar.org/paper/The-Secret-Sharer%3A-Evaluating-and-Testing-in-Neural-CarliniLiu/520ec00dc35475e0554dbb72f27bd2eeb6f4191d; Huseyin Inan, et al., Privacy Analysis in Language Models Via Training Data Leakage Report, arXiv (Feb. 22, 2021), https://arxiv.org/abs/2101.05405.

[29] See Yunhui Long, Understanding And Mitigating Privacy Risk In Machine Learning Systems, University of Illinois at Urbana-Champaign (2020), https://www.ideals.illinois.edu/bitstream/handle/2142/107972/LONGDISSERTATION-2020.pdf.

[30] Data should be appropriately biased (in a statistical sense) for what it's needed to do in order to have accurate predictions. However, beyond this, diverse concerns with unwanted bias exist, including factors that could make a system's outcomes morally or legally unfair. See Ninaresh Mehrabi et al., A Survey on Bias and Fairness in Machine Learning, USC Information Sciences Institute (Sept. 17, 2019), https://arxiv.org/pdf/1908.09635.pdf. For an illustration of ways fairness can be assessed across the AI lifecycle, see Sara Robinson, Building Machine Learning Models for Everyone: Understanding Fairness in Machine Learning, Google (Sept. 25, 2019), https://cloud.google.com/blog/products/ai-machinelearning/building-ml-models-for-everyone-understanding-fairness-in-machine-learning.

[31] For more examples on the law of armed conflict, see Artificial Intelligence and Machine Learning in Armed Conflict: A Human-Centered Approach, International Committee of the Red Cross (June 6, 2019), https://www.icrc.org/en/document/artificial-intelligence-and-machine-learning-armed-conflict-human-centredapproach.

B. **Representing Objectives and Tradeoffs.** Above, we described the goals of developing and fielding systems that align with key values through employing technologies, engineering efforts, and operational policies. Another important practice for aligning AI systems with values is to consider values as (1) embodied in choices about engineering tradeoffs and (2) as explicitly represented in the goals and utility functions of an AI system.[32]

On (1), multiple tradeoffs may be encountered with the engineering of an AI system. With AI, tradeoffs need to be made based on what is most valued (and the benefits and risks to those values)[33] including for high-stakes, high-risk pattern recognition, recommendation, and decision making under uncertainty. Decisions about tradeoffs for AI systems must be made about internal representations, policies of usage and controls, run-time execution monitoring, and thresholds. These include a number of well-known, inescapable engineering tradeoffs when it comes to building and using machine-learning to develop models for prediction, classification, and perception. For example, systems that perform recognition or prediction tasks can be set to work at different operating thresholds or settings (along a well characterized curve) where different settings change the tradeoff between precision and recall or the rates of true positives and false positives. By changing the settings, the ratio of true positives to false positives is changed. Often, one can raise the rate of true positives but will also raise the false negatives.[34] In high-stakes applications, different kinds of inaccuracies (e.g., missing a recognition and falsely recognizing) are associated with different outcomes and costs. For example, in a medical recommendation system, a false negative will lead to a missed or delayed treatment of an illness, while a false positive will lead to a potentially costly, but unnecessary and dangerous treatment for an illness that is not present. An engineer or policy maker can change the likelihood of each of these failures by shifting the threshold for an inferred probability of illness at which a recommendation for treatment is made. Further, investing greater resources in data and modeling will shift failure rates, and thus, frame additional questions about values around the engineering effort invested in systems employed in high-stakes settings. Thus, decisions about thresholds and about engineering investments, and understanding the influences of these decisions on the behavior of a system entail making value judgments. As with all engineering tradeoffs, making choices about tradeoffs explicitly and deliberately provides more transparency, accountability, and confidence in the process than making decisions implicitly and ad hoc as they arise.

---

[32] Mohsen Bayati, et al., Data-Driven Decisions for Reducing Readmissions for Heart Failure: General Methodology and Case Study, PLOS One Medicine (Oct. 2014), https://doi.org/10.1371/journal.pone.0109264; Eric Horvitz & Adam Seiver, Time-Critical Action: Representations and Application, Proceedings of the Thirteenth Conference on Uncertainty in Artificial Intelligence (Aug. 1997), https://arxiv.org/pdf/1302.1548.pdf.

[33] Jessica Cussins Newman, Decision Points in AI Governance: Three Case Studies Explore Efforts to Operationalize AI Principles, Berkeley Center for Long-Term Cybersecurity (May 5, 2020), https://cltc.berkeley.edu/ai-decision-points/ [hereinafter Newman, Decision Points in AI Governance].

[34] For more on the tradeoffs between false positive and false negative rates, and the implications of chosen thresholds, see Grother, Face Recognition Vendor Test (FRVT) Part Three: Demographic Effects.

On (2), systems may be guided by optimization processes that seek to maximize an objective function. [35] Such objectives can represent the desirability or the pursuit of a combination of independent goals. Various technical approaches (e.g., use of multi-attribute utility functions) may be employed to guide a system's actions based on an objective that is constructed by weighing several individual factors. In some cases, explicit weights are assigned to capture the asserted importance of each of the different factors. Sets of weightings on factors, and the inclusion versus exclusion of specific factors, can be viewed as embedding different values into a system. Here too, tradeoffs are made either explicitly or implicitly when setting different weights (of importance) to different objectives.[36] For example, there may be structural relationships among desired factors, such as the inverse relationship between the speed and safety at which an autonomous vehicle transports people.

Increasing the weighting of one desired factor (speed of travel) may necessarily reduce the weighting on another (safety of travel). As another example, when tuning a model for fairness, optimizing for one metric of fairness can cause a tradeoff in performance across the second metric.[37] As a result, it is important to acknowledge inherent tradeoffs and the need for setting or encoding values or preferences about tradeoffs, which requires someone or some organization to make a call about the trade.[38]

---

[35] Objective functions may also be referred to as utility functions or utility models.

[36] Optimal decisions may require making a decision when tradeoffs exist between two or more conflicting objectives. For example, a predictive maintenance system for aircraft will have objectives that are in tension including: minimizing false positives, minimizing false negatives, minimizing the need for instrumentation on the aircraft, maximizing the specificity of the recommended maintenance action, and adapting to new operational profiles the aircraft perform in over time.

[37] It is sometimes impossible to simultaneously satisfy different fairness criteria. See Yungfeng Zhang, et al., Joint Optimization of AI Fairness and Utility: A Human-Centered Approach, Association for Computing Machinery, AIES '20 (Feb. 7-8, 2020), https://dl.acm.org/doi/10.1145/3375627.3375862.

[38] See Analyses of Alternatives, Systems Engineering Guide, MITRE (May 2014), https://www.mitre.org/publications/systems-engineering-guide/acquisition-systems-engineering/acquisitionprogram-planning/performing-analyses-of-alternatives.

*Recommended Practices for Representing Objectives and Tradeoffs*

1. **Consider and document value considerations in AI systems and components by specifying how tradeoffs with accuracy are handled;** this includes selection of operating thresholds that have implications for performance, such as the precision (positive predictive value) and recall (sensitivity) of predictions or the true positive and false positive rates.[39]

2. **Consider and document value considerations in AI systems that rely on representations of objective functions,** especially when assigning weightings that capture the importance of different goals for the system.

3. **Conduct documentation, reviews, and set limits based on disallowed outcomes.** It is important to:

   - Be transparent[40] and keep documentation on assertions about the tradeoffs made, optimization justifications, and acceptable thresholds for false positives and false negatives.
   - During system development and testing, consider the potential need for context-specific changes in goals or objectives that would require a revision of parameters on settings or weightings on factors.
   - Establish explicit controls in specific use cases and have the capability to change or set controls, potentially by context or by policy, per organization.
   - Review documentation and run-time execution tradeoffs, potentially on a recurrent basis, by appropriate experts/authorities.
   - Acknowledge that performance characteristics are statistics over multiple cases, and that different settings and workloads have different performance.
   - Set logical limits based on disallowed outcomes, where needed, to put additional constraints on allowed performance.

---

[39] See Frank Liang, Evaluating the Performance of Machine Learning Models, Towards Data Science (Apr. 18, 2020), https://towardsdatascience.com/classifying-model-outcomes-true-false-positivesnegatives-177c1e702810.

[40] Throughout the Key Considerations, we refer to several types of documentation that should be conducted. To the extent feasible, this document should be shared publicly to engender the goals of transparency and public trust.

(4) Recommendations for Future Action

Future R&D is needed to advance capabilities for preserving and ensuring that developed or acquired AI systems will act in accordance with democratic values and the rule of law. For instance, there is a need for R&D to assure that the personal privacy of individuals is protected in the acquisition and use of data for AI system development.[41] This includes advancing ethical practices with the use of personal data, including disclosure and consent about data collection and use models (including uses of data to build base models that are later retrained and fine-tuned for specific tasks). R&D should also advance development of anonymity techniques and privacy-preserving technologies including homomorphic encryption and differential privacy techniques and identify optimal approaches for specific use cases. Research should focus upon advancing multi-party compute capabilities (to allow collaboration on the pooling of data from multiple organizations without sharing datasets), and developing a better understanding of the compatibility of the promising privacy preserving approaches with regulatory approaches such as the European Union's General Data Protection Regulation (GDPR), as both areas are important for allied cooperation.

## II. Engineering Practices

(1) Overview

The government, and its partners (including vendors), should adopt recommended practices for creating and maintaining trustworthy and robust AI systems that are auditable (able to be interrogated and yield information at each stage of the AI lifecycle to determine compliance with policy, standards, or regulations[42]); traceable (to understand the technology, development processes, and operational methods applicable to AI capabilities, e.g., with transparent and auditable methodologies, data sources, and design procedure and documentation[43]); interpretable (to understand the value and accuracy of system output[44]), and reliable (to perform in the intended manner within the intended domain of use[45]).

---

[41] In its Final Report, the Commission provides a fulsome assessment of where investment needs to be made; important R&D areas through the lens of ethics and responsible AI are integrated into R&D recommendations found in Chapters 3, 7, and 11. See Final Report, NSCAI (Mar. 1, 2021), https://www.nscai.gov/2021-finalreport/.

[42] See Inioluwa Deborah Raji, et al., Closing the AI Accountability Gap: Defining an End-to-End Framework for Internal Algorithmic Auditing, ACM FAT (Jan. 3, 2020), https://arxiv.org/abs/2001.00973 [hereinafter Raji, Closing the AI Accountability Gap].

There are no broadly directed best practices or standards (e.g., endorsed by the Secretary of Defense or Director of National Intelligence) in place to define how relevant organizations should build AI systems that are consistent with designated AI principles. But efforts in commercial, scientific, research, and policy communities are generating candidate approaches, minimal standards, and engineering proven practices to ensure the responsible design, development, and deployment of AI systems.[46]

While AI refers to a constellation of technologies, including logic-based systems, the rise in capabilities in AI systems over the last decade is largely attributable to capabilities provided by data-centric machine learning (ML) methods. New security and robustness challenges are linked to different phases of ML system construction and operations.[47] Several properties of the methods and models used in ML are associated with weaknesses that make the systems brittle and exploitable in specific ways—and vulnerable to failure modalities not seen in traditional software systems. Such failures can rise inadvertently or as the intended results of malicious attacks and manipulation. Attributes of machine learning training procedures and run-times linked to intentional and unintentional failures include: (1) the critical reliance on data for training, (2) the common use of such algorithmic procedures as differentiation and gradient descent to construct and optimize the performance of models, (3) the ability to probe models with multiple tasks or queries, and (4) the possibility of gaining access to information about models and their parameters.

Given the increasing consequences of failure in AI systems as they are integrated into critical uses, the various failure modes of AI systems have received significant attention. The exploration of AI failure modes has been divided into adversarial attacks[48] or unintended faults introduced throughout the lifecycle.[49] The pursuit of security and robustness of AI systems requires awareness, attention, and proven practices around intentional and unintentional failure modes.[50]

---

[43] Lopez, DOD Adopts 5 Principles.

[44] Model Interpretability in Azure Machine Learning (preview), Microsoft (Feb. 25, 2021), https://docs.microsoft.com/en-us/azure/machine-learning/how-to-machine-learning-interpretability.

[45] Lopez, DOD Adopts 5 Principles.

[46] See Newman, Decision Points in AI Governance; Raji, Closing the AI Accountability Gap; Miles Brundage, et al., Toward Trustworthy AI Development: Mechanisms for Supporting Verifiable Claims (Apr. 20, 2020), https://arxiv.org/abs/2004.07213 [hereinafter Brundage, Toward Trustworthy AI Development]; Saleema Amershi, et. al., Software Engineering for Machine Learning: A Case Study, Microsoft (Mar. 2019), https://www.microsoft.com/en-us/research/uploads/prod/2019/03/amershi-icse-2019_Software_Engineering_for_Machine_Learning.pdf [hereinafter Amershi, Software Engineering for Machine Learning].

[47] Elham Tabassi, et al., A Taxonomy and Terminology of 4 Adversarial Machine Learning (Draft NISTIR 8269), National Institute of Standards and Technology (Oct. 2019), https://nvlpubs.nist.gov/nistpubs/ir/2019/NIST.IR.8269-draft.pdf [hereinafter Tabassi, A Taxonomy and Terminology of 4 Adversarial Machine Learning (Draft NISTIR 8269)].

Intentional failures are the result of malicious actors explicitly attacking some aspect of (AI) system training or run-time behavior. Researchers and practitioners in the evolving area of Adversarial Machine Learning (AML) have created taxonomies of malicious attacks on machine learning training procedures and run-times. Attacks span ML training and testing and each has associated defenses.[51] Categories of intentional failures introduced by adversaries include training data poisoning attacks, model inversion, and ML supply chain attacks.[52] National security uses of AI are likely targets of sustained adversarial efforts; awareness of sets of potential vulnerabilities and proven practices for detecting attacks and protecting systems is critical. AI developed for this community must remain current with a rapidly developing understanding of the nature of vulnerabilities to attacks as these attacks grow in sophistication. Advances in new attack methods and vectors must be followed with care and recommended practices implemented around technical and process methods for mitigating vulnerabilities and detecting, alerting, and responding to attacks.

Unintentional failures can be introduced at multiple points in the AI development and deployment lifecycle. In addition to faults that can be inadvertently introduced into any software development effort (e.g., requirements ambiguity, coding errors, inadequate TEVV, flaws in tools used to develop and evaluate the system), distinct additional failure modes can be introduced for machine learning systems. Examples of unintentional AI failures (with particular relevance to deep learning and reinforcement learning) include reward hacking, side-effects, distributional shifts, and natural adversarial examples.[53] Another area of failure includes the inadequate specification of values per objectives represented in system utility functions (as described in Section 1 above on Representing Objectives and Trade-offs), leading to unexpected and costly behaviors and outcomes, akin to outcomes in the fable of the Sorcerer's Apprentice.[54] Additional classes of unintentional failures can arise as unexpected and potentially costly behaviors generated via the interactions of multiple distinct AI systems that are each developed and tested in isolation. The explicit or inadvertent composition of sets of AI systems within one's own services, forces, agencies, and between US systems and those of allies, adversaries, and potential adversaries, can lead to complex multi-agent situations with unexpected and poorly-characterized behaviors.[55]

---

[48] See Guofu Li, et al., Security Matters: A Survey on Adversarial Machine Learning, (Oct. 2018), https://arxiv.org/abs/1810.07339; Tabassi, A Taxonomy and Terminology of 4 Adversarial Machine Learning (Draft NISTIR 8269).

[49] See José Faria, Non-Determinism and Failure Modes in Machine Learning. 2017 IEEE 28th International Symposium on Software Reliability Engineering Workshops (Oct. 23-26, 2017), https://ieeexplore.ieee.org/document/8109300; Dario Amodei, et al., Concrete Problems in AI Safety, (July 25, 2016), https://arxiv.org/abs/1606.06565.

[50] Ram Shankar Siva Kumar, et al., Failure Modes in Machine Learning, (Nov. 11, 2019), https://docs.microsoft.com/en-us/security/engineering/failure-modes-in-machine-learning [hereinafter Kumar, Failure Modes in Machine Learning].

[51] See Tabassi, A Taxonomy and Terminology of 4 Adversarial Machine Learning (Draft NISTIR 8269).

[52] For 11 categories of attack, and associated overviews, see the "Intentionally-Motivated Failures Summary" in Kumar, Failure Modes in Machine Learning.

[53] Id.

## (2) Examples of Current Challenges

To make high-stakes decisions, and often in safety-critical contexts, Departments and agencies such as DoD and the IC must be able to depend on the integrity and security of the data that is used to train some kinds of ML systems. The challenges of doing so have been echoed by the leadership of the DoD and the Intelligence Community,[56] including concerns with detecting adversarial attacks such as data poisoning, sensor spoofing, and "enchanting attacks" (when the adversary lures a reinforcement learning agent to a designated target state that benefits the adversary).[57]

---

[54] Thomas Dietterich & Eric Horvitz, Rise of Concerns about AI: Reflections and Directions, Communications of the ACM, Vol. 58 No. 10 at 38-40 (Oct. 2015), http://erichorvitz.com/CACM_Oct_2015-VP.pdf.

[55] "Unexpected performance represents emergent runtime output, behavior, or effects at the system level, e.g., through unanticipated feature interaction, […] that was also not previously observed during model validation." See Colin Smith, et al., Hazard Contribution Modes of Machine Learning Components, AAAI-20 Workshop on Artificial Intelligence Safety (SafeAI 2020) at 4 (Feb. 7, 2020), https://ntrs.nasa.gov/archive/nasa/casi.ntrs.nasa.gov/20200001851.pdf.

[56] See Don Rassler, A View from the CT Foxhole Lieutenant General John N.T. "Jack" Shanahan, Director, Joint Artificial Intelligence Center, Department of Defense, Combating Terrorism Center at West Point (Dec. 2019), https://ctc.usma.edu/view-ct-foxhole-lieutenant-general-john-n-t-jack-shanahan-director-joint-artificialintelligence-center-department-defense/ ("I am very well aware of the power of information, for good and for bad. The profusion of relatively low-cost, leading-edge information-related capabilities and advancement of AIenabled technologies such as generative adversarial networks or GANs, has made it possible for almost anyone—from a state actor to a lone wolf terrorist—to use information as a precision weapon. What was viewed largely as an annoyance a few years ago has now become a serious threat to national security. Even more alarming, it's almost impossible to predict the exponential growth of these information-as-a-weapon capabilities over the next few years."); see also Dean Souleles, 2020 Spring Symposium: Building an AI Powered IC, Intelligence and National Security Alliance (Mar. 9, 2020), https://www.insaonline.org/2020spring-symposium-building-an-ai-powered-ic-event-recap/ ("We need to be thinking of authenticity of information and provenance of information.....How do you know that the news you are reading is authentic news? How do you know that its source has provenance? How can you trust the information of the world? And in this era of deep fakes and generative artificial neural networks scans that can produce images and texts and videos and audio that are increasingly indistinguishable from authentic, where then is the role of the intelligence officer? If you can no longer meaningfully distinguish truth from falsehood, how do you write an intelligence report?  How do you tell national leadership with confidence you believe something to be true or not to be true. That is a big challenge... We need systems that are reliable and understandable. We need to be investing in the gaps.").

[57] Naveed Akhtar & Ajmal Mian, Threat of Adversarial Attacks on Deep Learning in Computer Vision: A Survey (Feb. 26, 2018), https://arxiv.org/abs/1801.00553.

## (3) Recommended Practices

*Engineering Recommended Practices*

Critical engineering practices needed to operationalize AI principles (such as 'traceable' and 'reliable'[58]) are described in the non-exhaustive list below. These practices span design, development, and deployment of AI systems.

1. **Refine design and development requirements, informed by the concept of operations and risk assessment, including characterization of failure modes and associated impacts.** Conduct systems analysis of operations and identify mission success metrics. Identify potential functions that can be performed by AI technology. Incorporate early analyses of use cases and scenario development, assess general feasibility and compliance with disallowed outcomes expressed in policy. Make a critical assessment of the reproducibility and demonstrated technical maturity of specific candidate AI technologies. Reproducibility refers to how readily research results can be replicated by a third party, and is a significant concern in machine learning research.[59] The early analyses should include broad stakeholder engagement and hazard analyses, with domain experts and individuals with expertise and/or training in the responsible development and fielding of AI technologies. This requires for example asking key questions about potential disparate impact early in the development process and documenting deliberations, actions, and approaches used to ensure fairness and lack of unwanted bias in the machine learning application.[60] The feasibility of meeting these requirements may trigger a review of whether and where it is appropriate to use AI in the system being proposed. Opportunities exist to use experimentation, modeling/simulation, and rapid prototyping of AI systems to validate operational requirements and assess feasibility.[61]

---

[58] See Press Release, U.S. Department of Defense, DOD Adopts Ethical Principles for Artificial Intelligence (Feb. 24, 2020), https://www.defense.gov/Newsroom/Releases/Release/Article/2091996/dod-adopts-ethical-principles-for-artificial-intelligence/.

[59] Joelle Pineau, et al., Improving Reproducibility in Machine Learning Research (A Report from the NeurIPS 2019 Reproducibility Program), arXiv (Dec. 30, 2020), https://arxiv.org/abs/2003.12206.

[60] There is no single definition of fairness. System developers and organizations fielding applications must work with stakeholders to define fairness, and provide transparency via disclosure of assumed definitions of fairness. Definitions or assumptions about fairness and metrics for identifying fair inferences and allocations should be explicitly documented. This should be accompanied by a discussion of alternate definitions and rationales for the current choice. These elements should be documented internally as machine-learning components and larger systems are developed. This is especially important as establishing alignment on the metrics to use for assessing fairness encounters an added challenge when different cultural and policy norms are involved when collaborating on development and use with allies.

- **Risk assessment.** In conducting stakeholder engagement and hazard analysis, it is important to assess risks and tradeoffs with a diverse interdisciplinary group. This includes an analysis of the system's potential societal impact and the impacts of the system's failure modes. Prior to developing or acquiring a system, or conducting AI R&D in a novel area, risk assessment questions should be asked relevant to the national security context in critical areas, including questions about privacy and civil liberties, the law of armed conflict, human rights,[62] system security, and the risks of a new technology being leaked, stolen, or weaponized.[63]

2. **Produce documentation of the AI lifecycle:** Whether building and fielding an AI system or "infusing AI" into a preexisting system, require documentation[64] on:

---

[61] Design reviews take place at multiple stages in the Defense Acquisition process. Recent reforms to the Defense Acquisition System efforts, include the release of a new DoD 5000.02, which issues the "Adaptive Acquisition Framework" and an interim policy for a software acquisition pathway; this reflects efforts to further adapt the system to support agile and iterative approaches to software-intensive system development. See Software Acquisition, Defense Acquisition University (last visited June 18, 2020), https://aaf.dau.edu/aaf/software/; DoD Instruction 5000.02: Operation Of The Adaptive Acquisition Framework, Office of the Under Secretary of Defense for Acquisition and Sustainment (Jan. 23, 2020), https://www.esd.whs.mil/Portals/54/Documents/DD/issuances/dodi/500002p.pdf?ver=2020-01-23-144114-093.

[62] For more on the importance of human rights impact assessments of AI systems, see Report of the Special Rapporteur to the General Assembly on AI and its impact on freedom of opinion and expression, UN Human Rights Office of the High Commissioner (2018), https://www.ohchr.org/EN/Issues/FreedomOpinion/Pages/ReportGA73.aspx. For an example of a human rights risk assessment for AI in categories such as nondiscrimination and equality, political participation, privacy, and freedom of expression, see Mark Latonero, Governing Artificial Intelligence: Upholding Human Rights & Dignity, Data Society (Oct. 2018), https://datasociety.net/wp-content/uploads/2018/10/DataSociety_Governing_Artificial_Intelligence_Upholding_Human_Rights.pdf.

[63] For exemplary risk assessment questions that IARPA has used, see Richard Danzig, Technology Roulette: Managing Loss of Control as Many Militaries Pursue Technological Superiority, Center for a New American Security at 22 (June 28, 2018), https://s3.amazonaws.com/files.cnas.org/documents/CNASReport-Technology-Roulette-DoSproof2v2.pdf?mtime=20180628072101. Where appropriate, agencies should leverage standards and best practices from NIST's future risk management framework to mitigate identified privacy, civil liberties, and civil rights risks. See Pub. L. 116-283, sec 5301, William M. (Mac) Thornberry National Defense Authorization Act for Fiscal Year 2021, 134 Stat. 3388 (2021) (mandating NIST to establish a voluntary risk management framework, in consultation with industry, that will provide standards and best practices for assessing the trustworthiness of AI and mitigating risks from AI systems, including per aspects of privacy and fairness).

[64] Documentation recommendations build off of a legacy of robust documentation requirements. See Department of Defense Standard Practice: Documentation of Verification, Validation, and Accreditation (VV&A) For Models and Simulations, U.S. Department of Defense (Jan. 28, 2008), https://acqnotes.com/Attachments/MIL-STD-3022%20Documentation%20of%20VV&A%20for%20Modeling%20&%20Simulation%2028%20Jan%2008.pdf.

- If ML is used, the data used for training and testing, including clear and consistent annotation of data, the origin of the data (e.g., why, how, and from whom), provenance, intended uses, and any caveats with re-uses;[65]
- The algorithm(s) used to build models, characteristics about the model (e.g, training), and the intended uses of the AI capabilities separately or as part of another system;
- Connections between and dependencies within systems, and associated potential complications;
- The selected testing methodologies and performance indicators and results for models used in the AI component (e.g., confusion matrix and thresholds for true and false positives and true and false negatives area under the curve (AUC) as metrics for performance/error); this includes how tests were done, and the simulated or real-world data used in the tests—including caveats about the assumptions of the training and testing, per type of scenarios, per the data used in testing and training;
- Required maintenance, including re-testing requirements, and technical refresh. This includes requirements for re-testing, retraining, and tuning when a system is used in a different scenario or setting (including details about definitions of scenarios and settings) or if the AI system is capable of online learning or adaptation.

3. **Leverage infrastructure to support traceability, including auditability and forensics.**

Invest resources and build capabilities that support the traceability of AI systems. Traceability, critical for high-stakes systems, captures key information about the system development and deployment process for relevant personnel to adequately understand the technology.[66] It includes selecting, designing, and implementing measurement tools, logging, and monitoring and applies

---

to (1) development and testing of AI systems and components,[67] (2) operation of AI systems,[68] (3) users and their behaviors in engaging with AI systems or components,[69] and (4) auditing.[70] Audits should support analyses of specific actions as well as characterizations of longer-term performance. Audits should also be done to assure that performance on tests of the system and on real-world workloads meet requirements, such as fairness asserted at specification of the system and/or established by stakeholders.[71] When a criminal investigation requires it, forensic analyses of the AI system must be supported. A recommended practice is to carefully consider how you expose APIs for audit trails and traceability infrastructure in light of the potential vulnerability to an adversary detecting how an algorithm works and conducting an attack using counter AI exploitation.[72]

4. **For security and robustness, address intentional and unintentional failures.**

- **Adversarial attacks and use of robust ML methods.** Expand notions of adversarial attacks to include various "machine learning attacks," which may take the form of an attack through supply chain, online access, adversarial training data, or model inference attacks, including through Generative Adversarial Networks (GANS).[73] Agencies should seek latest

---

[67] Examples include logs of steps taking in problem and purpose definition, design, training and development. See e.g., Brundage, Toward Trustworthy AI Development.

[68] This includes logs of steps taken in operation which can support retrospective accident analysis. Id.

[69] Examples include logs of access and use of the system by operators, per understanding human access, oversight; nonrepudiation (e.g., cryptographic controls on access).

[70] Auditing examples include real-time system health and behavior monitoring, longer-term reporting, via logging of system recommendations, classifications, or actions and why they were taken per input, internal states of the system that were important in the chain of inferences and ultimate actions, and the actions taken, and logs to assure maintenance of accountability for decision systems (e.g. signoff for a specific piece of business logic).

[71] All of the above are consistent with, and support the fulfillment of, the DOD's AI Principle, Traceable. Documentation practices that support traceability (e.g. data sources and design procedures and documentation) are expanded upon in additional bullets throughout the Engineering Practices section. See Lopez, DOD Adopts 5 Principles ("Traceable: - The department's AI capabilities will be developed and deployed such that relevant personnel possess an appropriate understanding of the technology, development processes and operational methods applicable to AI capabilities, including with transparent and auditable methodologies, data sources and design procedures and documentation.").

[72] For example, "APIs are 'doors' to access digital infrastructures; thus, the security and resilience of digital environments will also depend on the robustness of the API infrastructure." Lorenzino Vaccari, et al., Application Programming Interfaces in Governments: Why, What and How, European Union Joint Research Centre at 13 (2020), .https://ec.europa.eu/jrc/en/publication/eur-scientific-and-technical-research-reports/application-programming-interfaces-governments-why-what-and-how.

[73] The approach is to simultaneously train two models: a generative model G that captures the data distribution, and a discriminative model D that estimates the probability that a sample came from the training data rather than G. As the generator gets better (producing ever more credible samples) the discriminator also improves (getting ever better at discerning real samples from the generated "fake"

technologies that demonstrate the ability to detect and notify operators of attacks, and also tolerate attacks.[74]

- **Follow and incorporate advances in intentional and unintentional ML failures.** Given the rapid evolution of the field of study of intentional and unintentional ML failures, national security organizations must follow and adapt to the latest knowledge about failures and proven practices for monitoring, detection, and engineering and run-time protections. Related efforts and R&D focus on developing and deploying robust AI methods.[75]

- **Adopt a DevSecOps lifecycle for AI systems** to include a focus on potential failure modes. This includes developing and regularly refining threat models to capture and consolidate the characteristics of various attacks in a way that can shape system development to mitigate vulnerabilities.[76] A matrixed focus for developing and refining threat models is valuable.

---

samples). This is useful for improving discriminator performance. Given the vulnerability of deep learning models to adversarial examples (slight changes in an input that produce significantly different results in output and can be used to confound a classifier), there has been interest in using adversarial inputs in a GAN framework to train the discriminator to better distinguish adversarial inputs. There is also considerable theoretical work being done on fundamental approaches to making DL more robust to adversarial examples. This remains an important focus of research. For more on adversarial attacks, see e.g., Ian Goodfellow et al., Generative Adversarial Networks, Universite de Montreal (June 10, 2014), https://arxiv.org/abs/1406.2661; Ian Goodfellow et. al., Explaining And Harnessing Adversarial Examples, Google (Mar. 20, 2015), https://arxiv.org/pdf/1412.6572.pdf; Kevin Eykholt, et al., Robust Physical-World Attacks on Deep Learning Visual Classification, Proceedings of the IEEE Conference on Computer Vision and Pattern Recognition at 1625–1634 (2018), https://arxiv.org/abs/1707.08945; Anish Athalye, et al., Synthesizing Robust Adversarial Examples, International conference on machine learning (2018), https://arxiv.org/pdf/1707.07397.pdf; Kevin Eykholt, et al., Physical Adversarial Examples for Object Detectors, USENIX Workshop on Offensive Technologies (2018), https://arxiv.org/abs/1807.07769; Yulong Cao, et al., Adversarial Sensor Attack on LiDAR-based Perception in Autonomous Driving, Proceedings of the 2019 ACM SIGSAC Conference on Computer and Communications Security (2019), https://dl.acm.org/doi/10.1145/3319535.3339815; Mahmood Sharif, et al., Accessorize to a Crime: Real and Stealthy Attacks on State-of-the-Art Face Recognition, Proceedings of the 2016 ACM SIGSAC Conference on Computer and Communications Security (2016) https://dl.acm.org/doi/10.1145/2976749.2978392; Stepan Komkov & Aleksandr Petiushko, Advhat: Real-World Adversarial Attack on Arcface Face ID System (Aug. 23, 2019), https://arxiv.org/abs/1908.08705.pdf. On directions with robustness, see e.g., Aleksander Madry, et al., Towards Deep Learning Models Resistant to Adversarial Attacks. MIT (Sept. 4, 2019), https://arxiv.org/abs/1706.06083 [hereinafter Madry, Toward Deep Learning Models Resistant to Adversarial Attacks]; Mathias Lecuyer, et al., Certified Robustness to Adversarial Examples with Differential Privacy, IEEE Symposium on Security and Privacy (2019), https://arxiv.org/abs/1802.03471; Eric Wong & J. Zico Kolter, Provable Defenses Against Adversarial Examples via the Convex Outer Adversarial Polytope, International Conference on Machine Learning (2018), https://arxiv.org/abs/1711.00851.

[74] Madry, Towards Deep Learning Models Resistant to Adversarial Attacks.

[75] See e.g., Id.; Thomas Dietterich, Steps Toward Robust Artificial Intelligence, AI Magazine at 3-24 (Fall 2017), https://www.aaai.org/ojs/index.php/aimagazine/article/view/2756/2644; Eric Horvitz, Reflections on Safety and Artificial Intelligence, Safe AI: Exploratory Technical Workshop on Safety and Control for AI (June 27, 2016), http://erichorvitz.com/OSTP-CMU_AI_Safety_framing_talk.pdf.

[76] See Andrew Marshall, et al, Threat Modeling AI/ML Systems and Dependencies, Microsoft (Nov. 11, 2019),

https://docs.microsoft.com/en-us/security/engineering/threat-modeling-aiml.

DevSecOps should address ML development, deployment, and when ML systems are under attack.[77]

- **Limit consequences of system failure through system architecture.** Build an overall system architecture that monitors component performance and handles errors when anomalies are detected; build AI components to be self-protecting and self-checking; and include aggressive stress testing under conditions of intended use. Where technically feasible, ensure that high consequence AI systems have overall system architectures that support robust recovery and repair or fail-fast and fail-over to a reliable degraded mode safe system.

5. **Conduct red teaming** for both intentional and unintentional failure modalities. Bring together multiple perspectives to rigorously challenge AI systems, exploring the risks, limitations, and vulnerabilities in the context in which they'll be deployed. Red teaming and general testing of the robustness of systems can be assisted by tools and practices, including the use of modeling and simulation and synthetic and augmented data.[78]

- To mitigate intentional failure modes—Assume an offensive posture that engages in creative and persistent attacks on systems and organizations, and defend against such attacks by employing methods that can make systems more resistant to adversarial attacks, work with adversarial testing tools, and deploy teams dedicated to trying to break systems and push them to violate rules for appropriate behavior.

- To mitigate unintentional failure modes—Test ML systems per a thorough list of realistic conditions they are expected to operate in. When selecting third-party components, consider the impact that a security vulnerability in them could have to the security of the larger system into which they are integrated. Have an accurate inventory of third-party components and a plan to respond when new vulnerabilities are discovered.[79]

---

[77] Ram Shankar Siva Kumar, et al., Adversarial Machine Learning—Industry Perspectives, 2020 IEEE Symposium on Security and Privacy (SP) Deep Learning and Security Workshop (Mar. 2021), https://arxiv.org/pdf/2002.05646.pdf.

[78] Synthetic data is generated according to some statistical profile for data features. See e.g., About, SyntheticMass (last accessed Apr. 6, 2021), https://synthea.mitre.org/about. Augmented data means taking existing training data and modifying it in a way that produces additional valid training data, e.g. flipping or cropping an image.

[79] See What are the Microsoft SDL Practices?, Microsoft (last accessed July 14, 2020), https://www.microsoft.com/en-us/securityengineering/sdl/practices.

- Because of the scarcity of required expertise and experience for AI red teams, organizations should consider establishing broader enterprise-wide communities of AI red teaming capabilities that could be applied to multiple AI developments (e.g., at a DoD service or IC element level, or higher).

**(4) Recommendations for Future Action**

- *For documentation:* There is an urgency for documentation strategy best practices.[80] Future work is needed to ensure sufficient documentation by all national security departments and agencies, including the precisions noted above in this section. In the meantime, national security departments and agencies should pilot documentation approaches across the AI lifecycle to help inform such a strategy.

- *To improve traceability:* While recommended practices exist for audit trails, standards have yet to be developed.[81] Future work is needed by standard setting bodies, alongside national security departments/agencies and the broader AI community (including industry), to develop audit trail requirements per mission needs for high-stakes AI systems including safety-critical applications.

- Future R&D is needed to advance capabilities for:

  o AI security and robustness—to cultivate more robust methods that can overcome adverse conditions; advance approaches that enable assessment of types and levels of vulnerability and immunity; and to enable systems to withstand or to degrade gracefully when targeted by a deliberate attack.

  o AI system risk assessment—to advance capabilities to support risk assessment including standard methods and metrics for evaluating degrees of auditability, traceability, interpretability, explainability, and reliability. For interpretability in particular, R&D is also needed to improve our understanding of the efficacy of interpretability tools and possible interfaces.

---

[80] See First Quarter Recommendations, NSCAI at 71-73 (Mar. 2020), https://www.nscai.gov/previous-reports/. Ongoing efforts to share best practices for documentation among government agencies through GSA's AI Community of Practice further indicate the ongoing need and desire for common guidance.

[81] For more on current gaps in audit trail standards for AI systems, see Brundage, Toward Trustworthy AI Development at 25 ("Existing standards often define in detail the required audit trails for specific applications. For example, IEC 61508 is a basic functional safety standard required by many industries, including nuclear power. Such standards are not yet established for AI systems.").

## III. System Performance

(1) Overview

Fielding AI systems in a responsible manner includes establishing confidence that the technology will perform as intended, especially in high-stakes scenarios.[82] An AI system's performance must be assessed,[83] including assessing its capabilities and blind spots with data representative of real-world scenarios or with simulations of realistic contexts,[84] and its reliability and robustness (i.e., resilience in real-world settings—including adversarial attacks on AI components) during development and in deployment.[85] For example, a system's performance on recognition tasks can be characterized by its false positives and false negatives on a test set representative of the environment in which a system will be deployed, and test sets can be varied in realistic ways to estimate robustness. Testing protocols and requirements are essential for measuring and reporting on system performance, including reliability, during the test phase (pre-deployment) and in operational settings. (The Commission uses industry terminology 'testing' to broadly refer to what the DoD calls "Test, Evaluation, Verification, and Validation" (TEVV) This testing includes both what DoD refers to as Developmental Test and Evaluation and Operational Test and Evaluation.). AI systems present new challenges to established testing protocols and requirements as they increase in complexity, particularly for operational testing. However, there are some existing methods to continuously monitor AI system performance. For example, high-fidelity performance traces and means for sensing shifts, such as distributional shifts in targeted scenarios, permit ongoing monitoring to ensure system performance does not stray outside of acceptable parameters; if inadequate performance is detected, they provide insight needed to improve and update systems.[86]

---

[82] This includes, for example, safety-critical scenarios or those where AI-assisted decision making would impact an individual's life or liberty.

[83] Ben Shneiderman, Human-Centered Artificial Intelligence: Reliable, Safe & Trustworthy, International Journal of Human–Computer Interaction (Mar. 23, 2020), https://doi.org/10.1080/10447318.2020.1741118 [hereinafter Shneiderman, Human Centered Artificial Intelligence: Reliable, Safe & Trustworthy].

[84] However, test protocols must acknowledge test sets may not be fully representative of real-world usage.

[85] See Brundage, Toward Trustworthy AI Development; Ece Kamar, et al., Combining Human and Machine Intelligence in Large-Scale Crowdsourcing, Proceedings of the 11th International Conference on Autonomous Agents and Multiagent Systems (June 2012), https://dl.acm.org/doi/10.5555/2343576.2343643 [hereinafter Kamar, Combining Human and Machine Intelligence in Large-Scale Crowdsourcing].

System performance characterization also includes assessing robustness. As noted above, this entails determining how resilient the system is in real-world settings where there may be blocking and handling of attacks and where natural real-world variation exists.[87] In addition to reliability, robustness, and security, system performance must also measure compliance with requirements derived from values such as fairness.

When evaluating system performance, it is especially important to take into account holistic, end-to-end system behavior. Emergence is the principle that entities exhibit properties which are meaningful only when attributed to the whole, not to its parts. Emergent system behavior can be viewed as a consequence of the interactions and relationships among system elements rather than the independent behavior of individual elements. It emerges from a combination of the behavior and properties of the system elements and the system's structure or allowable interactions between the elements, and may be triggered or influenced by a stimulus from the system's environment.[88]

The System Engineering Community and the National Security Community have focused on system of systems engineering for years,[89] but AI-intensive systems introduce additional opportunities and challenges for emergent performance. Given the requirement to establish and preserve justified confidence in the performance of AI systems, attention must be paid to the potential for undesired interactions and emergent performance as AI systems are composed. This composition may include pipelines where the output of one system is part of the input for another in a potentially complex and distributed ad hoc pipeline.[90] As a recent study of the software engineering challenges introduced by developing and deploying AI systems at scale notes, "AI components are more difficult to handle as distinct modules than traditional software components—models may be 'entangled' in complex ways."[91] These challenges are pronounced when the entanglement is the result of system composition and integration.

---

[86] For a technical paper that puts monitoring in development lifecycle context, see Amershi, Software Engineering for Machine Learning. For a good example of open source frameworks to support, see Overview, Prometheus (last accessed June 18, 2020), https://prometheus.io/docs/introduction/overview/.

[87] Joel Lehman, Evolutionary Computation and AI Safety: Research Problems Impeding Routine and Safe Real-world Application of Evolution (Oct. 4, 2019), https://arxiv.org/abs/1906.10189 [hereinafter Lehman, Evolutionary Computation and AI Safety].

[88] Greg Zacharias, Autonomous Horizons: The Way Forward, Air University Press at 61 (Mar. 2019), https://www.airuniversity.af.edu/Portals/10/AUPress/Books/b_0155_zacharias_autonomous_horizons.pdf.

[89] Judith Dahmann & Kristen Baldwin, Understanding the Current State of US Defense Systems of Systems and the Implications for Systems Engineering, Presented at IEEE Systems Conference (Apr. 7-10, 2008), https://ieeexplore.ieee.org/document/4518994.

[90] D. Sculley, et al., Machine Learning: The High Interest Credit Card of Technical Debt, Google (2014), https://research.google/pubs/pub43146/ [hereinafter Sculley, Machine Learning: The High Interest Credit Card of Technical Debt].

As America's AI-intensive systems may increasingly be composed (including through ad hoc opportunities to integrate systems) with allied AI-intensive systems, this becomes a topic for coordination with allies as well. Multi-agent systems are being explored and adopted in multiple domains,[92] as are swarms, fleets, and teams of autonomous systems.[93]

## (2) Examples of Current Challenges

Having justified confidence in AI systems requires assurances that they will perform as intended, including when interacting with humans and other systems. Testing directed at providing these assurances will increasingly encounter challenges when compared with testing of traditional software systems:

- With respect to the U.S. government, although recent extensions to acquisition processes[94] are intended to accommodate rapid iterative development for software intensive systems, there are still challenges in integrating the agile development process typical of machine learning into enterprise processes. The data that shapes supervised learning is as important as code, requiring version control and configuration management for volumes of data as well as code for retest and regression testing. Training a model is processor-intensive and time-consuming, a challenge for rapid build and deploy cycles.

- There is an ongoing need for common infrastructure for developing and testing AI systems, including common frameworks/architectures, common built-in instrumentation for transparency and interpretability in testing and operation, and common testbeds and test ranges.

---

[91] Amershi, Software Engineering for Machine Learning (illustrating non-monotonic error as a possible complexity result from model entanglement).

[92] Ali Dorri, et al., Multi-Agent Systems: A Survey, IEEE Access at 28573-28593 (June 19, 2018), https://ieeexplore.ieee.org/stamp/stamp.jsp?tp=&arnumber=8352646.

[93] Andrew Ilachinski, AI, Robots, and Swarms: Issues, Questions, and Recommended Studies, CNA (Jan. 2017), https://www.cna.org/CNA_files/PDF/DRM-2017-U-014796-Final.pdf.

[94] For extensions to the DoD acquisition process, see DoD Instruction 5000.82: Acquisition of Information Technology (IT), U.S. Department of Defense (Apr. 21, 2020), https://www.esd.whs.mil/Portals/54/Documents/DD/issuances/dodi/500082p.pdf?ver=2020-04-21-153621-140.

To minimize performance problems and unanticipated outcomes, testing is essential. Yet, there is a lack of common metrics to assess trustworthiness that AI systems will perform as intended.

(3) Recommended Practices

Critical practices for ensuring optimal system performance are described in the following non-exhaustive list:

A. **Training and Testing: Procedures should cover key aspects of performance and appropriate performance metrics.**

1. **Use regularly updated standards for testing and reporting of system performance.** Standards for metrics and reporting are needed to adequately:

    a. Achieve consistency across testing and test reporting for critical areas.

    b. Test for blind spots as a specific failure mode of importance to some ML implementations.[95]

    c. Test for fairness. When testing for fairness, sustained fairness assessments are needed throughout development and deployment, including assessing a system's accuracy and errors relative to one or more agreed to statistical definitions of fairness[96] and documenting deliberations made on the appropriate fairness metrics to use.[97] Agencies should also conduct outcome and impact analysis to detect when subtle assumptions in

---

[95] Ramya Ramakrishnan, et al., Blind Spot Detection for Safe Sim-to-Real Transfer, Journal of Artificial Intelligence Research 67 at 191-234 (2020), https://www.jair.org/index.php/jair/article/view/11436.

[96] There is no single definition of fairness. System developers and organizations fielding applications must work with stakeholders to define fairness, and provide transparency via disclosure of assumed definitions of fairness. Definitions or assumptions about fairness and metrics for identifying fair inferences and allocations should be explicitly documented. This should be accompanied by a discussion of alternate definitions and rationales for the current choice. These elements should be documented internally as machine-learning components and larger systems are developed. This is especially important as establishing alignment on the metrics to use for assessing fairness encounters an added challenge when different cultural and policy norms are involved when collaborating on development and use with allies.

[97] Examples of tools available to assist in assessing and mitigating bias in systems relying on machine learning include Aequitas by the University of Chicago, Fairlearn by Microsoft, AI Fairness 360 by IBM, and PAIR and ML-fairness-gym by Google.

the system concept of operations and requirements are showing up as unexpected and undesired outcomes in the operational environment.[98]

d.  Articulate system performance. This includes ways to communicate to the end user the meaning/significance of performance metrics, e.g., through a probability assessment, based on sensitivity and specificity. It also requires clear documentation of system performance (across diverse environments or contexts), including information content of model output.

2.  **Consider and document the representativeness of the data and model for the specific context at hand.** For machine learning models, challenges exist when transferring a model to a context/setting that differs from the one for which it was trained and tested. When using classification and prediction technologies, challenges with representativeness of data used in analyses, and fairness/accuracy of inferences and recommendations made with systems leveraging that data when applied in different populations/contexts, should be considered explicitly and documented. As appropriate, robust and reliable methods can be used to enable model generalization and transfer beyond the training context.

3.  **Evaluate an AI system's performance relative to current benchmarks where possible.** Benchmarks should assist in determining if an AI system's performance meets or exceeds current best performance.

4.  **Evaluate aggregate performance of human-machine teams.** Consider that the current benchmark might be the current best performance of a human operator or the composed performance of the human-machine team. Where humans and machines interact, it is important to measure the aggregate performance of the team rather than the AI system alone.[99]

---

[98] See Microsoft's AI Fairness checklist as an example of an industry tool to support fairness assessments, Michael A. Madaio et al., Co-Designing Checklists to Understand Organizational Challenges and Opportunities around Fairness in AI, CHI 2020 (Apr. 25-30, 2020), http://www.jennwv.com/papers/checklists.pdf [hereinafter Madaio, Co-Designing Checklists to Understand Organizational Challenges and Opportunities around Fairness in AI].

[99] Kamar, Combining Human and Machine Intelligence in Large-scale Crowdsourcing.

5. **Provide sustained attention to reliability and robustness:** Various kinds of AI systems often demonstrate impressive performance on average but can fail in ways that are unexpected in any specific instance. The performance potential of an AI system is often roughly determined by experiment and test, rather than by any predictive analytics. AI can have blinds spots and unknown fragilities.[100] Focus on tools and techniques to carefully bound assumptions of robustness of the AI component in the larger system architecture, and provide sustained attention to characterizing the actual performance envelope for nominal and off-nominal conditions throughout development and deployment.[101] For systems of particularly high potential consequences of failure, considerable architecture and design work will have been put into making the overall system fail-safe. Special attention must be paid to confirming that error detection and recovery or fail-over mechanisms in the system are effective (e.g., by design reviews, testing that challenges boundary conditions and assumptions, and instrumentation and monitoring).

6. **For systems of systems, test machine-machine/multi-agent interaction.** Individual AI systems will be combined in various ways in an enterprise to accomplish broader missions beyond the scope of any single system. For example, pipelines of AI systems will exist where the output of one system serves as the input for another AI system. (The output of a track management and classifier system might be input to a target prioritization system which might in turn provide input to a weapon/target pairing tool.) Multiple relatively independent AI systems can be viewed as distinct agents interacting in the environment of the system of systems, and some of these agents will be humans in and on the loop. Industry has encountered and documented problems in building 'systems of systems' out of multiple AI systems.[102] A related problem is poor backward compatibility when the performance of one model in a pipeline is enhanced and may result in degrading the overall system of system behavior.[103] These problems in composition illustrate emergent performance, as described in the conceptual overview portion of this section. Unexpected failures that transpire in systems of systems may not be the result of any one component failing but may instead be based in the interactions among the composed systems.[104]

---

[100] John Launchbury, A DARPA Perspective on Artificial Intelligence, DARPA, (last accessed June 18, 2020), https://www.darpa.mil/about-us/darpa-perspective-on-ai (noting that machine learning is "statistically impressive, but individually unreliable").

[101] Shneiderman, Human-Centered Artificial Intelligence: Reliable, Safe & Trustworthy.

[102] One example is "Hidden Feedback Loops", where systems that learn from external world behavior may also shape the behavior they are monitoring. See Sculley, Machine Learning: The High Interest Credit Card of Technical Debt. See also Cynthia Dwork, et al., Individual Fairness in Pipelines, arXiv (Apr. 12, 2020), https://arxiv.org/abs/2004.05167; Megha Srivastava, et al., An Empirical Analysis of Backward Compatibility in Machine Learning Systems, KDD '20: Proceedings of the 26th ACM SIGKDD International Conference on Knowledge Discovery & Data Mining (Aug. 2020), https://dl.acm.org/doi/10.1145/3394486.3403379 [hereinafter Srivastava, An Empirical Analysis of Backward Compatibility in Machine Learning Systems].

A frequent cause of failures in composed systems is the violation of assumptions that were not previously challenged; therefore, a priority during testing should be to challenge ("stress test") interfaces and usage patterns with boundary conditions and challenges to assumptions about the operational environment and use. This is focused on both unintended violations of assumptions from system composition and also deliberate challenges to the system by adversarial attacks.

B. **Maintenance and deployment.** Given the dynamic nature of AI systems, recommended practices for maintenance are also critically important. These include:

1. **Specify maintenance requirements** for datasets as well as for systems, given that their performance can degrade over time.[105]

2. **Continuously monitor and evaluate AI system performance**, including the use of high-fidelity traces to determine continuously if a system is going outside of acceptable parameters (including operational performance measures and established constraints for fairness and core values), both during pre-deployment and operation.[106] This includes measuring system

---

[103] Srivastava, An Empirical Analysis of Backward Compatibility in Machine Learning Systems.

[104] D. Sculley, et al., Hidden Technical Debt in Machine Learning Systems, NIPS '15: Proceedings of the 28th International Conference on Neural Information Processing Systems (Dec. 2015), https://dl.acm.org/doi/10.5555/2969442.2969519.

[105] Artificial Intelligence (AI) Playbook for the U.S. Federal Government, Artificial Intelligence Working Group, ACT-IAC Emerging Technology Community of Interest, (Jan. 22, 2020),

https://www.actiac.org/act-iac-white-paper-artificial-intelligence-playbook.

[106] Beyond accuracy, high-fidelity traces capture other parameters of interest/musts, including fairness, fragility (e.g. whether a system degrades gracefully versus unexpectedly fails), security/attack resilience, and privacy leakage. Often instrumentation results from execution are treated as time-series data and can be analyzed by a variety of anomaly detection techniques to identify unexpected or changing characteristics of system performance. See Meir Toledano, et al., Real-Time Anomaly Detection System for Time Series at Scale, KDD 2017: Workshop on Anomaly Detection in Finance (2017), http://proceedings.mlr.press/v71/toledano18a/toledano18a.pdf. DoD recently updated its acquisition processes to improve "the ability to deliver warfighting capability at the speed of relevance" See DoD 5000 Series Acquisition Policy Transformation Handbook, U.S. Department of Defense (Jan. 15, 2020),

performance per acceptable parameters in terms of both reliability and values.[107] It also includes assessing statistical results for performance over time, for example, to detect emergent bias or anomalies.[108] As with any instrumentation and storage or monitoring design (e.g., flight data recorders, system logs), design tradeoffs must be made between the potential value of the data captured and monitored and the costs for bandwidth, computing, and storage imposed.

3. **Conduct iterative and sustained testing and validation.** Be wary that training and testing that provide characteristics on capabilities might not transfer or generalize to specific settings of usage (for example lighting conditions in some applications may be very different for scene interpretation); thus, testing and validation may need to be done recurrently, and at strategic intervention points, but especially for new deployments and classes of task.[109]

4. **Monitor and mitigate emergent behavior.** There will be instances where systems are composed in ways not anticipated by the developers (e.g., opportunistic integration with an ally's system). These use cases clearly can't be adequately addressed at development time; some aspects of confidence in the composition must be shifted to monitoring the actual performance of the composed system and its components. For emergent performance

---

https://www.acq.osd.mil/ae/assets/docs/DoD%205000%20Series%20Handbook%20(15Jan2020).pdf. These include revised policies for acquiring software-intensive systems and components. Relevant here, program managers are now required to "ensure that software teams use iterative and incremental software development methodologies," and use modern technologies "to achieve automated testing, continuous integration and continuous delivery of user capabilities, frequent user feedback/engagement (at every iteration if possible), security and authorization processes, and continuous runtime monitoring of operational software" Ellen Lord, Software Acquisition Pathway Interim Policy and Procedures, Memorandum from the Undersecretary of Defense, to Joint Chiefs of Staff and Department of Defense Staff (Jan. 3, 2020), https://www.acq.osd.mil/ae/assets/docs/USA002825-19%20Signed%20Memo%20(Software).pdf. See also Ori Cohen, Monitor! Stop Being A Blind Data-Scientist (Oct. 8, 2019), https://towardsdatascience.com/monitor-stop-being-a-blind-data-scientist-ac915286075f; Mace, Pivot Tracing.

[107] Values parameters could include pre-determined thresholds for acceptable false positive or false negative rates for fairness, or parameters set regarding data or model leakage in the context of privacy.

[108] Lehman, Evolutionary Computation and AI Safety.

[109] Eric Breck, et al., The ML Test Score: A Rubric for ML Production Readiness and Technical Debt Reduction, 2017 IEEE International Conference on Big Data, (Dec. 11-14, 2017), https://ieeexplore.ieee.org/stamp/stamp.jsp?arnumber=8258038&tag=1.

concerns when AI systems are composed, there are advances in runtime assurance/verification[110] and feature interaction management[111] that can be adapted.

**(4) Recommendations for Future Action**

- Future R&D is needed to advance capabilities for:

  o Testing, Evaluation, Verification, and Validation (TEVV) of AI systems - to develop a better understanding of how to conduct persistent TEVV of AI systems throughout an iterative development and deployment lifecycle and build checks and balances and ability to perform real-time updates into an AI system. Includes complex system testing—to increase our understanding of and ability to have confidence in emergent performance of composed AI systems. Improved methods are needed to understand, predict, and control systems-of-systems so that when AI systems interact with each other, their interaction does not lead to unexpected negative outcomes.

  o Multi-agent scenario understanding—to advance the understanding of interacting AI systems, including the application of game theory to varied and complex scenarios, and interactions between cohorts composed of a mixture of humans and AI technologies.

- Basic definitional work has been ongoing for years on how to characterize key properties such as fairness and explainability. Progress on a common understanding of the concepts and requirements is critical for progress in widely used metrics for performance.

- Significant, ongoing work is needed to establish what appropriate metrics should be used to assess system performance across attributes for responsible AI according to

---

application/context profiles. (Such attributes, for example, include fairness, interpretability, reliability and robustness.) Future work is needed to develop: (1) definitions, taxonomy, and metrics needed to enable agencies to better assess AI performance and vulnerabilities, and; (2) metrics and benchmarks to assess reliability and intelligibility of produced model explanations. In the near term, guidance is needed on: (1) standards for testing intentional and unintentional failure modes; (2) exemplar datasets for benchmarking and evaluation, including robustness testing and red teaming, and; (3) defining characteristics of AI data quality and training environment fidelity (to support adequate performance and governance).[112]

- International collaboration and cooperation is needed to:

    o Align on how to test and verify AI system reliability and performance along shared values (such as fairness and privacy). Establishing how to test systems will include measures of performance based on common standards, and may have implications for the types of traceability that will need to be incorporated into system design and development. Such collaboration on common testing for reliability and adherence to values will be critical among allies and partners to enable interoperability and trust. Additionally, these efforts could potentially include dialogues between the United States and strategic competitors regarding establishing common standards of AI safety and reliability testing in order to reduce the chances of inadvertent escalation.[113]

---

[112] The 2021 NDAA expansion of the National Institute of Standards & Technology (NIST) mission authorizes the standards body to provide such guidance. See Pub. L. 116-283, sec. 5301, William M. (Mac) Thornberry National Defense Authorization Act for Fiscal Year 2021, 134 Stat. 3388 (2021); see also Elham Tabassi, NIST AI Update, NIST at 14 (Mar. 4, 2021), https://csrc.nist.gov/CSRC/media/Presentations/ai-and-ndaa-requirements-nsc-ai-commission-report/images-media/AI%20and%20NDAA%20Requirements_%20NSC%20AI%20Commission%20Report%20Tabassi.pdf (noting that Title LIII, Sec. 5301 expands NIST mission to include "advancing collaborative frameworks, standards, guidelines for AI, supporting the development of a risk-mitigation framework for AI systems, and supporting the development of technical standards and guidelines to promote trustworthy AI systems.").

[113] For research regarding common interests in ensuring safety-critical systems work as intended (e.g. in a reliable manner) to avoid destabilization/escalatory dynamics, see Andrew Imbrie & Elsa Kania, AI Safety, Security, and Stability Among Great Powers: Options, Challenges, and Lessons Learned for Pragmatic Engagement, CSET, (Dec. 2019), https://cset.georgetown.edu/research/ai-safety-security-and-stability-among-great-powers-options-challenges-and-lessons-learned-for-pragmatic-engagement/.

## IV. Human-AI Interaction & Teaming

(1) Overview

Responsible AI development and fielding requires striking the right balance of leveraging human and AI reasoning, recommendation, and decision-making processes. Ultimately, all AI systems will have some degree of human-AI interaction as they will all be developed to support humans. In some settings, the best outcomes will be achieved when AI is designed to augment human intellect, or to support human-AI collaboration more generally. In other settings, however, time-criticality and the nature of tasks may make some aspects of human-AI interaction difficult or suboptimal.[114] Where the human role is critical in real-time decisions because it is more appropriate, valuable, or designated as such by our values, AI should be intentionally designed to effectively augment and support human understanding, decision making, and intellect. Sustained attention must be focused on optimizing the desired human-machine interaction throughout the AI system lifecycle. It is important to think through the use criteria that are most relevant depending on the model. Models are different for human-assisted AI decision-making, AI-assisted human decision-making, pure AI decision-making, and AI-assisted machine decision-making.

(2) Examples of Current Challenges

There is an opportunity to develop AI systems to complement and augment human understanding, decision making, and capabilities. Decisions about developing and fielding AI systems aimed at specific domains or scenarios should consider the relative strengths of AI capabilities and human intellect across expected distributions of tasks, considering AI system maturity or capability and how people and machines might coordinate.

Designs and methods for human-AI interaction can be employed to enhance human-AI teaming.[115] Methods in support of effective human-AI interaction can help AI systems to understand when and how to engage humans for assistance, when AI systems should take initiative to assist human operators, and, more generally, how to support the creation of effective human-AI teams. In engaging with end users, it may be important for AI systems to infer and share with end users well-calibrated levels of confidence about their inferences, so as to provide human operators with an ability to weigh the importance of machine output or pause to consider details behind a recommendation more carefully. Methods, representations, and machinery can be employed to provide insight about AI inferences, including the use of interpretable machine learning.[116] Research directions include developing and fielding machinery aimed at reasoning about human strengths and weaknesses, such as recognizing and responding to the potential for costly human biases of judgment and decision making in specific settings.[117] Other work centers on mechanisms that consider the ideal mix of initiatives, including when and how to rely on human expertise versus on AI inferences.[118] As part of effective teaming, AI systems can be endowed with the ability to detect the focus of attention, workload, and interruptability of human operators and consider these inferences in decisions about when and how to engage with the operators.[119] Directions of effort include developing mechanisms for identifying the most relevant information or inferences to provide end users of different skills in different settings.[120] Consideration must be given to the prospect introducing bias, including potential biases that may arise because of the configuration and sequencing of rendered data. For example, IC research[121] shows that confirmation bias can be triggered by the order in which information is displayed, and this order can consequently impact or sway intel analyst decisions. Careful design and study can help to identify and mitigate such bias.

---

[114] The need for striking the right balance of human involvement in situations of time criticality is not unique to AI. For instance, DoD systems dating back to the 80s have been designed to react to airborne threats at speeds faster than a human would be capable of. See MK 15 - Phalanx Close-In Weapons System (CIWS), U.S. Navy (Jan. 15, 2019), https://www.navy.mil/Resources/Fact-Files/Display-FactFiles/Article/2167831/mk-15-phalanx-close-in-weapon-system-ciws/

[115] Saleema Amershi, et al., Guidelines for Human-AI Interaction, Proceedings of the CHI Conference on Human Factors in Computing Systems (May 2019), https://dl.acm.org/doi/10.1145/3290605.3300233.

---

[116] Rich Caruana, et al., Intelligible Models for HealthCare: Predicting Pneumonia Risk and Hospital 30-day Readmission, Proceedings of the 21th ACM SIGKDD International Conference on Knowledge Discovery and Data Mining (Aug. 2015), https://www.semanticscholar.org/paper/Intelligible-Models-for-HealthCare%3A-Predicting-Risk-Caruana-Lou/cb030975a3dbcdf52a01cbd1c140711332313e13.

[117] Eric Horvitz, Reflections on Challenges and Promises of Mixed-Initiative Interaction, AAAI Magazine 28 Special Issue on Mixed-Initiative Assistants (2007), http://erichorvitz.com/mixed_initiative_reflections.pdf.

[118] Eric Horvitz, Principles of Mixed-Initiative User Interfaces, Proceedings of CHI '99 ACM SIGCHI Conference on Human Factors in Computing Systems (May 1999), https://dl.acm.org/doi/10.1145/302979.303030; Kamar, Combining Human and Machine Intelligence in Large-scale Crowdsourcing.

[119] Eric Horvitz, et al., Models of Attention in Computing and Communications: From Principles to Applications, Communications of the ACM 46(3) at 52-59 (Mar. 2003), https://cacm.acm.org/magazines/2003/3/6879-models-of-attention-in-computing-and-communication/fulltext.

[120] Eric Horvitz & Matthew Barry, Display of Information for Time-Critical Decision Making, Proceedings of the Eleventh Conference on Uncertainty in Artificial Intelligence (Aug. 1995), https://arxiv.org/pdf/1302.4959.pdf.

[121] There has been considerable research in the IC on the challenges of confirmation bias for analysts. Some experiments demonstrated a strong effect that the sequence in which information is presented alone can shape analyst interpretations and hypotheses. Brant Cheikes, et al., Confirmation Bias in Complex Analyses, MITRE (Oct. 2004), https://www.mitre.org/sites/default/files/pdf/04_0985.pdf. This highlights the care that is required when designing the human machine teaming when complex, critical, and potentially ambiguous information is presented to analysts and decision makers.

**(3) Recommended Practices**

Critical practices to ensure optimal human-AI interaction are described in the non-exhaustive list below. These recommended practices span the entire AI lifecycle.

*Human-AI Interaction Recommended Practices*

A. **Identification of functions of human in design, engineering, and fielding of AI**

1. **Given AI and human capabilities and complementarities, as well as requirements for accountability and human judgment, define the tasks of humans and the goals and mission of the human-machine team across the AI lifecycle.** This entails noting needs for feedback loops, including opportunities for oversight.

2. **Define functions and responsibilities of humans during system operation and assign them to specific individuals.** Functions will vary for each domain and each project within a domain; they should be periodically revisited as model maturity and human expertise evolve over time.

B. **Explicit support of human-AI interaction and collaboration**

1. **Extend Human-AI design methodologies and guidelines.**

   - **Design methodologies.** Develop methodologies that improve understanding of human-AI interaction and provide specific guidance and requirements that can be assessed.

   - **Design guidelines.** AI systems designs should take into consideration the defined tasks of humans in human-AI collaborations in different scenarios; ensure the mix of human-machine actions in the aggregate is consistent with the intended behavior, and accounting for the ways that human and machine behavior can co-evolve;[122] and also avoid automation bias (e.g., placing unjustified confidence in the results of the computation) and unjustified reliance on humans in the loop as failsafe mechanisms. Allow for auditing of the human-AI pair, not only the AI in isolation, which could be a secondary expert examining a subset of cases. Designs should be transparent (e.g., about why and how a system did what it did, system updates, or new capabilities) so that there is an understanding the AI is working day-to-day and to allow for an audit trail if things go wrong.[123] Based on context and mission need, designs should ensure usability of AI systems by AI experts, domain experts, and novices, as appropriate.[124] Both transparency and usability will depend on the audience.

---

[122] Patricia L. McDermott et al., Human-machine Teaming Systems Engineering Guide, MITRE (Dec. 2018), https://www.mitre.org/publications/technical-papers/human-machine-teaming-systems-engineering-guide; Shneiderman, Human-Centered Artificial Intelligence: Reliable, Safe & Trustworthy.

2. **Employ algorithms and functions in support of interpretability and explanation.** Algorithms and functions that provide individuals with task-relevant knowledge and understanding need to take into consideration that key factors in an AI system's inferences and actions can be understood differently by various audiences. These audiences span real-time operators who need to understand inferences and recommendations for decision support, engineers and data scientists involved in developing and debugging systems, and other stakeholders including those involved in oversight. Interpretability and explainability exists in degrees; what's needed in terms of explainability will depend on who is receiving the explanation, what the context is, and the amount of time available to deliver and process this explanation. In this regard, interpretability intersects with traceability, audit, and documentation practices.

3. **Design systems to provide cues to the human operator(s) about the level of confidence the system has in the results or behaviors of the system.**[125] AI system designs should appropriately convey uncertainty and error bounding. For instance, a user interface should convey system self-assessment of confidence alerts when the operational environment is significantly different from the environment the system was trained for, and indicate internal inconsistencies that call for caution.

---

4. **Refine policies for machine-human handoff and control of initiative.** Policies, and aspects of human computer interaction, system interface, and operational design, should define when and how information or tasks should be handed off from a machine to a human operator and vice versa. Include checks to continually evaluate whether distribution of tasks is working. Special attention should be given to the fact that humans may freeze during an unexpected handoff due to the processing time the brain needs, potential distractions, or the condition during which the handoff occurs. The same may be true with an AI system which may not fully understand the human's intent during the handoff and may consequently make unexpected actions.

5. **Leveraging traceability to assist with system development and understanding.** Traceability processes must capture details about human-AI interactions to retroactively understand where challenges occurred, and why, in order to improve systems and their use in the future and for redress. Infrastructure and instrumentation[126] can also help assess humans, systems, and environments to gauge the impact of AI at all levels of system maturity; and to measure the effectiveness and performance for hybrid human-AI systems in a mission context.

6. **Conduct training.** Train and educate individuals responsible for AI development and fielding, including human operators, decision makers, and procurement officers. Training should help the workforce better interact, collaborate with, and be supported by AI systems, including understanding AI tools. Training also should include experiences with use of systems in realistic situations. Beyond training in the specifics of the system and application, operators of systems with AI components, especially systems that perform classification or pattern recognition, should receive education that includes fundamentals of AI and data science, including coverage of key descriptors of performance, including rates of false negatives and false positives, precision and recall, and sensitivity and specificity.

- **Periodic certification and refresh.** In addition to initial programs of training, operators should receive ongoing refresher trainings. Beyond being scheduled periodically, refresher trainings are appropriate when systems are deployed in new settings and unfamiliar scenarios. Refresh on training is also needed when predictive models are revised with new or additional data as the performance of systems may shift with such updates introducing behaviors that are unfamiliar to human operators.[127]

---

[126] Infrastructure includes tools (hardware and software) in the test environment that support monitoring system performance (such as the timing of exchanges among systems, or the ability to generate test data). Instrumentation refers to the presence of monitoring and additional interfaces to provide insight into a specific system under test.

## (4) Recommendations for Future Action

Future R&D is needed to advance capabilities for:

- Enhanced human-AI interaction:

  - To progress the ability of AI technologies to perceive and understand the meaning of human communication, including spoken speech, written text, and gestures. This research should account for varying languages and cultures, with special attention to diversity given that AI typically performs worse in cases with gender and racial minorities.

  - To improve human-machine teaming. This should include disciplines and technologies centered on decision sciences, control theory, psychology, economics (human aspects and incentives), and human factors engineering, such as human-AI interfaces, to enhance situational awareness and make it easier for users to do their work. Human-AI interaction and the mechanisms and interfaces that support such interactions, including richer human-AI collaborations, will depend upon mission needs and appropriate degrees of autonomy versus human oversight and control. R&D for human-machine teaming should also focus on helping systems understand human blind spots and biases, and optimizing factors such as human attention, human workload, ideal mixing of human and machine initiatives, and passing control between the human and machine. For effective passing of control, and to have effective and trusted teaming, R&D should further enable humans and machines to better understand intent and context of handoff.

  - To advance human-AI and AI-AI teaming. R&D is needed to optimize the ability of humans and AI to work together to undertake complex, evolving tasks in a variety of environments, as well as for diverse groupings of machines, such as autonomous drones, to cooperate with each other, broader systems, and human counterparts to achieve shared objectives.

---

[127] Gagan Bansal, et al., Updates in Human-AI Teams: Understanding and Addressing the Performance/Compatibility Tradeoff, AAAI (July 7, 2019), https://doi.org/10.1609/aaai.v33i01.33012429.

- Ongoing work is needed to train the workforce that will interact with, collaborate with, and be supported by AI systems. In its First Quarter Recommendations, the Commission provided recommendations for such training.[128]

  o **Workforce training.** A complementary best practice for Human-AI Interaction is training the workforce to understand tools they're using; as AI gets democratized, it will also get misused. For probabilistic systems, concepts and ideas that are important in system operation should be understood; for operators this includes understanding concepts such as precision, recall, sensitivity and specificity, and ensuring operators know how to interpret the confidence in inferences that well-calibrated systems convey.

## V. Accountability and Governance

**(1)  Overview**

National security departments and agencies must specify who will be held accountable for both specific system outcomes and general system maintenance and auditing, in what way, and for what purpose. Government must address the difficulties in preserving human accountability, including for end users, developers, testers, and the organizations employing AI systems. End users and those ultimately affected by the actions of an AI system should be offered the opportunity to appeal an AI system's determinations. And, finally, accountability and appellate processes must exist not only for AI decisions, but also for AI system inferences, recommendations, and actions.

**(2)  Examples of Current Challenges**

Overseeing entities must have the technological capacity to understand what in the AI system caused the contentious outcome. For example, if a soldier uses an AI-enabled weapon and the result violates international law of war standards, an investigating body or military tribunal should be able to re-create what happened through auditing trails and other documentation. Without policies requiring such technology and the enforcement of those policies, proper accountability would be elusive if not impossible. Moreover, auditing trails and documentation will prove critical as courts begin to grapple with whether AI system's determinations reach the requisite standards to be admitted as evidence.[129] Building the traceability infrastructure to permit auditing (as described in the Engineering Practices section) will increase the costs of building AI systems and take significant work—a necessary investment given our commitment to accountability, discoverability, and legal compliance.

---

[128] See First Quarter Recommendations, NSCAI at 69-70 (Mar. 2020), https://www.nscai.gov/previous-reports/.

**(3) Recommended Practices**

Critical accountability and governance practices are identified in the non-exhaustive list below.

*Accountability and Governance Recommended Practices*

1. **Appoint full-time responsible AI leads to join senior leadership.** Every department and agency critical to national security and each branch of the armed services, at a minimum, should have a dedicated, full-time responsible AI lead who is part of the senior leadership team. Such leads should oversee the implementation of the Key Considerations recommended practices alongside the department/agency's respective AI principles. This includes: driving responsible AI training; serving as subject matter experts regarding existing and proposed responsible AI policy and best practices; leading interagency best practice sharing for responsible AI; and shaping procurement policy and guidance for product managers to ensure alignment with recommended practices and adopted AI principles. The department's responsible AI lead should determine the responsible AI governance structure to ensure centralized and consistent policies are applied across the department, and internally coordinate across the department to ensure synergistic implementation of Responsible AI policies and programs.

2. **Identify responsible actors.** Determine and document who is accountable for a specific AI system or any given part of an AI system and the processes involved with it. This includes identifying who is responsible for the development or procurement; operation (including the system's inferences, recommendations, and actions during usage) and maintenance of an AI system; as well as the authorization of a system and enforcement of policies for use. Determine and document the mechanism/structure for holding such actors accountable and to whom should that mechanism/structure be disclosed to ensure proper oversight.

---

[129] For more on the difficulties of admitting ML evidence, see Patrick Nutter, Machine Learning Evidence: Admissibility and Weight, University of Pennsylvania Journal of Constitutional Law (2019), https://scholarship.law.upenn.edu/jcl/vol21/iss3/8/.

3. **Require technology to strengthen accountability processes and goals.** Document the chains of custody and command involved in developing and fielding AI systems. Policy should establish clear requirements about information that should be captured about the development process (via traceability) and about system performance and behavior in operation (run-time monitoring) to support reliability and robustness as well as auditing for oversight; these are further tailored by AI policy leads and engineers. These requirements should be tailored for the specific constraints imposed by policy (e.g., governing data retention) and technical issues (e.g., capacity of bandwidth, compute, and storage). This will allow the government to know who was responsible at which point in time. Improving traceability and auditability capabilities will allow agencies to better track a system's performance and outcomes.[130]

4. **Adopt policies to strengthen accountability and governance.** Identify or, if lacking, establish policies that allow individuals to raise concerns about irresponsible AI development/fielding, e.g. via an ombudsman. This requires ensuring a governance structure is in place to address grievances and harms if systems fail, which supports feedback loops and oversight to ensure that systems operate as they should. Agencies should institute specific oversight and enforcement practices, including: auditing and reporting requirements; a mechanism that would allow thorough review of the most sensitive/high-risk AI systems to ensure auditability and compliance with responsible use and fielding requirements; an appealable process for those found at fault of developing or using AI irresponsibly; and grievance processes for those affected by the actions of AI systems. Agencies should leverage best practices from academia and industry for conducting internal audits and assessments,[131] while also acknowledging the benefits offered by external audits.[132]

---

[130] See Raji, Closing the AI Accountability Gap.

[131] See Raji, Closing the AI Accountability Gap ("In this paper, we present internal algorithmic audits as a mechanism to check that the engineering processes involved in AI system creation and deployment meet declared ethical expectations and standards, such as organizational AI principles"); see also Madaio, Co-Designing Checklists to Understand Organizational Challenges and Opportunities Around Fairness in AI.

[132] For more on the benefits of external audits, see Brundage, Toward Trustworthy AI Development. For an agency example, see Aaron Boyd, CBP Is Upgrading to a New Facial Recognition Algorithm in March, Nextgov.com (Feb. 7, 2020), https://www.nextgov.com/emerging-tech/2020/02/cbp-upgrading-new-facial-recognition-algorithm-march/162959/ (highlighting a NIST algorithmic assessment on behalf of U.S. Customs and Border Protection).

5. **Support external oversight.** Remain responsive and facilitate Congressional oversight through documentation processes and other policy decisions.[133] For instance, supporting traceability and specifically documentation to audit trails, will allow for external oversight.[134] Internal self-assessment alone might prove to be inadequate in all scenarios.[135] Congress can provide a key oversight function throughout the AI lifecycle, asking critical questions of agency leadership and those responsible for AI systems.

(4) Recommendations for Future Action

Currently no external oversight mechanism exists specific to AI in national security. Notwithstanding the important work of Inspectors General in conducting internal oversight, open questions remain as to how to complement current practices and structures.

## Acknowledgments

Useful feedback on the Key Considerations manuscript was provided by Anne Bowser, Erin Hahn, and David Danks. Special thanks to Caroline Danauy for legal analysis and editorial review. We also thank Lance Lantier for insights on DoD policies and directives, and Nik Marda, Samuel Trotter, and Jaide Tarwid for editorial support.

---

[133] Maranke Wieringa, What to Account for When Accounting for Algorithms, Proceedings of the 2020 ACM FAT Conference, (Jan. 2020), https://dl.acm.org/doi/10.1145/3351095.3372833.

[134] Raji, Closing the AI Accountability Gap.

[135] Brundage, Toward Trustworthy AI Development